\documentclass{article}

\usepackage{arxiv}

\usepackage[utf8]{inputenc} 
\usepackage[T1]{fontenc}    
\usepackage{hyperref}       
\usepackage{url}            
\usepackage{booktabs}       
\usepackage{amsfonts}       
\usepackage{nicefrac}       
\usepackage{microtype}      
\usepackage{lipsum}
\usepackage{graphicx}
\usepackage{authblk}
\graphicspath{ {./images/} }

\title{Impact of the COVID-19 outbreak on the Italian
Twitter vaccination debate: A network–based
analysis}

\author[1]{Veronica Lachi}
\author[1]{Giovanna Maria Dimitri}
\author[2]{Alessandro Di Stefano}
\author[3]{Pietro Li{\`o}}
\author[1,*]{Monica Bianchini}
\author[1,*]{Chiara Mocenni}

\affil[1]{University of Siena, Siena, Italy}
\affil[2]{University of Teesside, United Kingdom}
\affil[3]{University of Cambridge, Cambridge, United Kingdom}

\affil[*]{Equal contribution}

\begin{document}
\maketitle

\begin{abstract}
Vaccine hesitancy, or the reluctance to be vaccinated, is a phenomenon that has recently become particularly significant, in conjunction with the vaccination campaign against COVID-19. During the lockdown period, necessary to control the spread of the virus, social networks have played an important role in the
Italian debate on vaccination, generally representing the easiest and safest way to exchange opinions
and maintain some form of sociability. Among social network platforms, Twitter has assumed a strategic role in driving the public opinion, creating compact groups of users sharing similar views towards the utility, uselessness or even dangerousness of vaccines. In this paper, we present a new, publicly available, dataset of Italian tweets, TwitterVax, collected in the period January 2019--May 2022. Considering monthly data, gathered into forty--one retweet networks --- where nodes identify users and edges are present between users who have retweeted each other ---, we performed community detection within the networks, analyzing their evolution and polarization with respect to NoVax and ProVax users through time. This allowed us to clearly discover debate trends as well as identify potential key moments and actors in opinion flows,
characterizing the main features and tweeting behavior of the two communities.
\end{abstract}


\section{Introduction}
Vaccines are one of the most powerful weapons to fight infectious diseases. The use of vaccines has, in fact, helped to drastically reduce epidemic mortality rates in the 20th century \cite{centers1999impact}. 
Nonetheless, we can observe the presence of an ongoing phenomenon, called \textit{vaccine hesitancy}, which can be identified as a delay in acceptance or a clear refusal of vaccination, despite the availability of vaccination services \cite{schuster2015review}. Indeed, vaccination skepticism is a phenomenon that has existed since the first vaccine became available. However, vaccine hesitancy is currently a growing global attitude --- supported and amplified by the ease of finding controversial information on the Internet --- that can pose a problem in avoiding the outbreak of communicable diseases. 
Therefore, recognizing the importance of the phenomenon and the high risk to which it exposes the world population, especially in disadvantaged areas, is essential \cite{Simas21}. For this purpose, the Strategic Advisory Group of Experts (SAGE) on Immunization of the World Health Organization (WHO) has established, since 2012, a specific working group on the subject, led by a joint WHO/Unicef Secretariat group.\\
The reasons that lead to hesitancy or refusal have been widely studied in the last few decades. For example, in \cite{bertoncello2020socioeconomic}, it is shown that increasing levels of perceived economic hardship are associated with vaccine hesitancy, just as less parental education is significantly associated with vaccine refusal for children. However, not only socio--economic conditions are linked to vaccine hesitancy, which is a multifaceted phenomenon where cognitive, psychological and cultural factors all play a critical role \cite{browne2015going, hornsey2018psychological,murphy2021psychological, pomares2020association}. \\
Recently, due to the outbreak of the COVID-19 pandemic, the issue of vaccine hesitancy has become particularly important. Despite overwhelming evidence showcasing the effectiveness of vaccines \cite{kim2021looking}, surpassing even alternative measures like contact tracing and lockdowns\cite{cencetti2020using,alfano2020efficacy}, a large swath of the population has continued to refuse inoculation, endangering public health and economic and social life \cite{sallam2021covid}.
In this specific situation, in fact, social opinions on COVID-19 vaccines were negatively affected also by concerns related to the unprecedented speed with which they were developed \cite{machingaidze2021understanding}, with sometimes confusing media communications and unregulated social media information sources fomenting such anti--vaccination sentiment --- not to mention conspiracy theories, which include population control through 5G technology or the extermination of humanity through vaccines \cite{betsch2012opportunities,  burki2019vaccine,kata2010postmodern}. This phenomenon has been exasperated by the presence of so--called \textit{echo chambers}, virtual environments where like--minded people reinforce their opinions through repeated interactions that amplify their political leanings, beliefs and attitudes \cite{cota2019quantifying,del2016spreading,garimella2017effect,garimella2018political,o2015echo}. Such a mechanism can lead to increasingly polarized debates and phenomena of extremism \cite{moussaid2015amplification}. To get an idea of the diffusion of certain forms of thinking, the Annual Report 2021 by Censis (Centro Studi Investimenti Sociali, https://www.censis.it/rapporto-annuale-censis) reveals that: \textit{''For 5.9\% of Italians (about 3 million) Covid does not exist, for 10,9\% the vaccine is useless, for 31.4\% it is an experimental drug and people who get vaccinated \textit{act as guinea pigs}, for 12.7\% science produces more harm than good [$\dots$], for 19.9\% 5G is a sophisticated tool to monitor people''}. \\
Similarly to many other political and social issues, social networks represent a natural source of aggregation and a powerful medium for the largely uncontrolled dissemination of information.Undoubtedly, social media has had a considerable impact
on the health sector, as the user
sentiment can be used to understand collective panic or for the dissemination of reliable and unreliable medical claims
 \cite{hernandez2018social}. Additionally, social media has a recognized role in healthcare campaigns, both targeted at businesses and by government agencies and non-profits to combat rumors, encourage behavioral change, and share information, enabling the audience to engage and share feedback.
Specifically, Twitter is one of the most used social media \cite{auxier2021social} and can also be identified as a popular source of health information. For these reasons, it can provide realistic insights on the society perception regarding vaccination. Twitter administration itself explicitly tried to prevent the dissemination of misleading information regarding COVID-19, publishing a series of rules for users (https://help.twitter.com/it/rules-and-policies/medical-misinformation-policy), yet leaving to their common sense the respect of the code of ethics on the matter. \\
The aim of this work is that of exploiting Twitter data to study the evolution over the past three years of the Italian vaccine debate since, recently, vaccines have become a very political issue in Italy \cite{brandmayr2021public}. The contribution of this paper is manifold. First, we show that, with the outbreak of COVID-19, the vaccine debate has changed: it has significantly intensified and, furthermore, the discussion
has gone from being widespread among all interested users (mostly parents of preschoolers) to being concentrated in the hands of a few influential hubs. Second, we have detected and monitored the NoVax and ProVax communities. The relative proportion of the two communities has not changed significantly over time and ProVax users are the most numerous but also the least active. Moreover, we have identified core NoVax users as well as core ProVax users, demonstrating that the former outnumber the latter  --- who have, anyway, more followers, mostly among verified users. 
Finally, we provide a new dataset of Italian tweets, TwitterVax, collected between January 2019 and May 2022, which is publicly available at https://github.com/veronicalachi/TwitterVax.\footnote{{Despite being in contrast with the Twitter ToS, we made the decision to publicly share the tweet texts. We did so because we firmly believe that this act brings significant benefits to the research community.}}\\
The paper is organized as follows. 
In Section \ref{sec:RelatedWorks}, we present an overview of related works in the context of vaccine debates and social network analysis. In Section \ref{sec:methods}, we describe our approach to data collection and analysis, while in Section \ref{sec:res}
we present and discuss the obtained results. Finally, Section \ref{sec:concl} collects some conclusions and outlines future research perspectives.

\section{Related works}
\label{sec:RelatedWorks}
Several works in the literature --- many of which make use of network--based approaches --- investigate the structure and characteristics of the vaccination debate. \\
In \cite{bello2017detecting}, an analysis of Twitter data and official vaccination coverage rates showed that vaccine opinions from Twitter users could affect the vaccination decision--making process; moreover, the application of a community detection algorithm led to the identification of two user communities: one in support of vaccination, including important and influential users, and one against vaccination, characterized by a lower level of interaction. Using a structural network approach, in \cite{cossard2020falling}, it was shown that vaccination skeptics, as well as advocates, reside in their own distinct echo chambers and that the structure of these two communities is different, with skeptics organized into highly connected clusters and supporters characterized by the presence of influential hub users. A social network approach was also used in \cite{milani2020visual} to demonstrate that NoVax users retweeted frequently each other while ProVax users formed a fragmented network, with fewer connections; moreover, while the ProVaxes were mostly healthcare workers, the NoVaxes were mainly parents and activists. In \cite{lutkenhaus2019mapping}, seven different communities were identified, including health workers, writers and journalists, anti--establishment people and international vaccination advocates; contents shared by the healthcare community hardly reached other communities, while messages tweeted by anti--establishment users managed to filter to other communities.\\
If several works have been proposed studying the structure of the vaccine debate and the interactions between vaccination advocates and skeptics, still little research has been conducted on how the outbreak of the COVID-19 pandemic has influenced this controversial discussion. In \cite{bonnevie2021quantifying}, a comparison of tweets from the four months prior to the onset of COVID-19 and tweets from the four months following the outbreak of the pandemic reveals that vaccine opponents on Twitter increased by 80\%. 
Finally, in \cite{Naseem21}, the COVID-19 debate is analyzed by focusing on people who interacted and shared opinions on Twitter. Based on a sentiment dataset, called COVIDSenti, composed of 90 000 COVID-19--related tweets collected in the early stages of the pandemic (from February to March 2020), it was shown that, at the onset of the pandemic,  people favored lockdown while, as expected, sentiment shifted by mid--March.  \\
As far as the Italian scenario is concerned, so far few works have been published. In \cite{gori2021mis}, a new collected dataset was used to describe the polarization and volumes of tweets, only focusing on a few months just after the vaccination campaigns (between October 2020 and January 2021). The dataset was crawled using only the four words "vaccino/i" and "vaccinazione/i" and manually annotated, showing a higher percentage of NoVax users. Furthermore, in \cite{bellodi2022comparing}, sentiment analysis was proposed for COVID-19 vaccine hesitancy posts, collected from several different social networks, with an attention towards the COVID-19 vaccine or specifically the booster shot. In \cite{lenti2022global}, more than 316 million Twitter messages were gathered between October 2019 and March 2021 to identify the amount of disinformation flows among users of different countries. Indeed, the tweets were written in more than eighteen languages (including Italian), with a focus on the global debate rather than considering national realities. Finally, in \cite{crupi2022echoes}, a comprehensive study was conducted by collecting a dataset of 16,223,749 Italian tweets spanning from September 2019 to December 2021. The main objective was to investigate the impact of unprecedented experiences and measures related to the COVID-19 pandemic on polarization in vaccine discussions. The results obtained clarified that, despite the outbreak of COVID-19, the echo chamber phenomenon within vaccine discussions persisted. \\
The aim of the present work is to extend the research on the Italian vaccination discussion, particularly characterizing the structural evolution of the debate and investigating the changes in the NoVax and ProVax communities in terms of size, productivity, and core users.

\section{Materials \& methods}
\label{sec:methods}
Twitter, a widely recognized social networking platform, had its data publicly accessible via its application programming interface (API) until February 2023.
In this paper, we use a network--based approach to study the evolution of Italian public opinion on vaccines over time. Transforming the unstructured information available from Twitter into graph data has allowed us not only to employ common metrics used on networks to estimate their topological features but also to apply some powerful algorithms for community detection.
\subsection{Data Collection}
\label{subsec:DataCollection}
\begin{figure}[!t]
\centering
\includegraphics[width=2.0in]{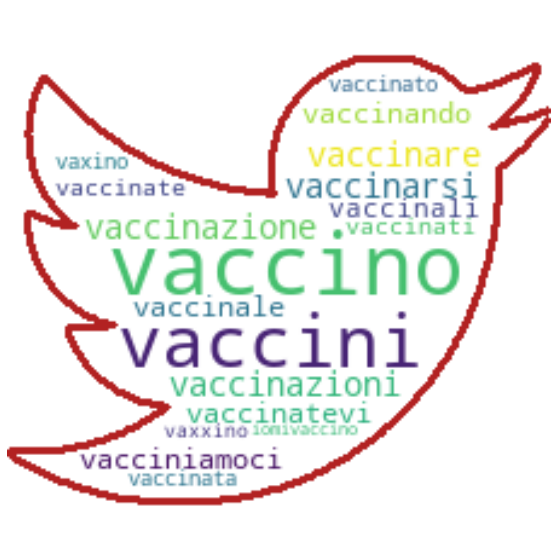}%
\caption{Keywords used in the Twitter API calls. The font size identifies the number of times the various words occur in our dataset.}
\label{fig:key}
\end{figure}

To perform data collection, the Twitter API v2 has been employed. In particular, the Twitter tweets/search/all endpoint\footnote{https://developer.twitter.com/en/docs/twitter-api/tweets/search/api-reference/get-tweets-search-all}  has been used, which allows us to search the full archive of tweets and to filter them using a set of keywords.
By crawling the Twitter archive based on twenty Italian vaccine--related terms as keywords, we were able to collect 9,068,389 tweets --- contributing to the vaccination debate --- posted by 300,653 users from 1 January 2019 to 31 May 2022. 
In particular, the twenty keywords selected to build the TwitterVax dataset were: \textit{vaccino, vaccini, vaccinazione, vaccinazioni, vaccinare, vaccinarsi, vaccinatevi, vacciniamoci, vaccinando, vaccinale, vaccinali, vaccinati, vaccinate, vaccinata, vaccinato, va@@ino, va..ino, vaxino, vaxxino, \#iomiovaccino}. Some of the words, \textit{vaxxino} for example, were specific jargon related to the debate and were frequently used in tweets by both the ProVax and NoVax communities. In contrast to the approach taken in \cite{crupi2022echoes}, we have made the decision to exclusively employ Italian keywords in our analysis. In Figure \ref{fig:key}, the most common words in our dataset are shown, where the font size is a quantitative indicator of the amount of occurrences of a specific word in the TwitterVax collection.
Moreover, we further filtered the tweets to return Italian texts only. 
Finally, we decided to study the dynamics of users’ opinions about vaccines by dividing the entire period into forty--one sub--periods, of the duration of one month each. The one--month sampling resulted heuristically the one that best preserves the information on the evolution of the community structure. \\
Interestingly, when using the Search API, some content is omitted from the dataset, especially content that has been deleted or posted by suspended users. The findings of \cite{pfeffer2022sample} shed light on this issue, revealing that a considerable proportion (72\%) of tweets captured through the Streaming Search API at the time of posting were not retrieved via the Historical Search API. This stark contrast in retrieval rates serves as a strong indication that a significant amount of Twitter content is swiftly removed from the platform, either due to user deletions or Twitter's enforcement actions against violations of community norms. This phenomenon highlights the dynamic nature of the platform, wherein a considerable portion of tweets is ephemeral, quickly disappearing from public view.
\subsection{Network construction and indicators}
\label{subsec:NetworkConstruction}
Using the data from each month, we constructed an ordered set of forty--one networks. In each graph, nodes represent the active users of the platform in that period, while undirected edges connect users that have retweeted each other at least once. This approach addresses a limitation highlighted in \cite{crupi2022echoes}, where an edge was considered only if there were at least two retweets, resulting in the exclusion of over half of the initial users. 
Since retweeting is the act of sharing another user's post without modifying it, people tend to retweet content they approve and to build a relationship with users with the same opinion about certain topics \cite{barbera2015tweeting, garimella2018quantifying,metaxas2015retweets}. Therefore, relationships between users induce an “agreement network”, in which groups of people with similar leaning are strongly connected to each other. 
\\
In order to reduce the dimension of the networks, tough keeping the most significant information, only the largest connected components have been used. In this way, the change in the discussion volume can be observed and the evolution of the network structure --- showing how the debate evolves in time --- can be described, by measuring its connectivity features. In particular, the following metrics were considered in the analysis.
\begin{itemize}
    \item \textit{Density} --- It represents the proportion of possible relationships in the network that are actually present \cite{van2010graph} or, in other words, it is calculated as the ratio of the number of edges in a graph over the number of edges of the complete graph with the same number of nodes. Therefore, it provides a measure of how dense a graph is in terms of edge connectivity.
    \item \textit{Average clustering coefficient} --- The local clustering coefficient of a node in a graph quantifies how close its neighbours are to being a clique (i.e., a complete graph). In other words, when computed on a single node, the clustering coefficient is a measure of how complete the neighborhood of a node is. Therefore, the overall level of clustering in a network is measured by averaging the clustering coefficient \cite{van2010graph,watts1998collective} over all the network nodes.The clustering coefficient of a single node is calculated as:
    $$
    c_v=\frac{2T(v)}{deg(v)(deg(v)-1)}
    $$
    where $T(v)$ is the the number of connected triangles including node $v$ and $deg(v)$ is the degree of $v$. 
    \item \textit{s--metric} --- The s--metric \cite{li2005towards} is a structural metric that provides a measure of the extent to which a graph is scale--free. It is calculated as:
    $$
    s(\mathcal{G})=\sum_{(v,u)\in \mathcal{E}} deg(v)\times deg(u)  
    $$
    where $\mathcal{G}=(\mathcal{V},\mathcal{E})$ is a graph, $\mathcal{V}$ is the set of its vertices or nodes, $\mathcal{E}$ is the set of its edges, and $deg(v)$ is the degree of the node $v$.
    A scale--free network \cite{barabasi1999emergence} is a network whose degree distribution follows a power law, at least asymptotically. The main characteristic of scale--free networks is the presence of nodes with a degree that significantly exceeds the average degree. 
\end{itemize}

\subsection{Community detection and user labeling}
\label{subsec:communityDetection}
Since the vaccination discussion is highly contentious, it is reasonable to assume that, at different stages, each active user takes a specific position in the debate. In particular, given the high polarization of this topic \cite{schmidt2018polarization}, we hypothesize that just two cohorts of people populate the platform: the vaccination skeptics (NoVax) and the vaccination advocates (ProVax). Classifying each user into these two categories for each month is not an easy task. The simplest and most reliable way to label a user as NoVax or ProVax is to analyze the content of their tweets, which is impossible to do manually, due to the large amount of text. 
Furthermore, despite significant advancements in natural language processing (NLP) methods for analyzing tweets, also for COVID-19 contents \cite{muller2020covid,nguyen2022emotion}, this area of research continues to present ongoing challenges. The brevity of tweets, constrained to a maximum of 280 characters, coupled with their complex and unusual semantics, poses significant obstacles. As a result, the effective utilization of NLP techniques in this domain remains a subject of active investigation. 
Thus, we decided to use the structure of the network to infer the opinion of the users. \\
Identifying communities of highly connected users within a network is one of the most popular and challenging problems in network science \cite{fortunato2016community, liu2020deep}. 
To perform community detection we have used the multi--level graph partitioning algorithm METIS \cite{karypis1998fast}. 
The choice of METIS was driven by several reasons. Firstly, the extensive usage of METIS in significant studies within the literature \cite{garimella2018quantifying,cossard2020falling} ensures the comparability of our results with existing research. Additionally, METIS has proven its effectiveness in handling retweet networks \cite{garimella2018quantifying}. Furthermore, the partitions generated by METIS consistently exhibit superior quality, with improvements ranging from 0\% to 50\% compared to spectral partitioning algorithms \cite{barnard1994fast}. Indeed, extensive experimentation across diverse graphs has demonstrated METIS's superiority in terms of speed, surpassing other widely utilized partitioning algorithms by one to two orders of magnitude \cite{karypis1997metis}. Finally, a clustering algorithm was recommended in which the number of communities was defined in advance, as we aimed to specifically identify two communities (ProVaxes and NoVaxes). \\
The idea behind METIS is to create successively smaller graphs $\mathcal{G}_1$, $\mathcal{G}_2$, $\ldots$, $\mathcal{G}_k$ from $\mathcal{G}_0=(\mathcal{V},\mathcal{E})$, to obtain a partition of $\mathcal{G}_k$ in a short time, and project the partition back onto $\mathcal{G}_0$, while refining it at each step.
In particular, METIS consists of two stages: coarsening and refinement.
\begin{enumerate}
\item \textit{Coarsening} --- The original graph, $\mathcal{G}_0$, is transformed into sequentially smaller graphs $\mathcal{G}_1$, $\mathcal{G}_2$, $\ldots$, $\mathcal{G}_k$, such that $|\mathcal{V}_0| > |\mathcal{V}_1| > |\mathcal{V}_2| > \ldots > |\mathcal{V}_k|$. If $\mathcal{G}_k$ is meant to be a good representation of $\mathcal{G}_0$, a good partitioning of $\mathcal{G}_k$ represents a fairly good partitioning of $\mathcal{G}_0$. $\mathcal{G}_k$ is small enough to make partitioning very quick. 
\item \textit{Refinement} --- The partition $\mathcal{P}_k$ is projected back onto $\mathcal{P}_{k-1}$, $\ldots$, $\mathcal{P}_0$. After each projection $\mathcal{P}_i$, the partitioning is refined using a greedy algorithm. Each partition $\mathcal{P}_i$, for $0 \le i \le k-1$, is refined before projecting to $\mathcal{P}_{i-1}$.
\end{enumerate}
Following the procedure adopted in \cite{cossard2020falling}, we have applied METIS repeatedly one hundred times, choosing two as the number of communities. This leads to a vector of one hundred elements, with entries equal to 0 or 1, representing the partition assignment for each user. Averaging the assigned partition across the vector allows to obtain a score between 0 and 1, which represents the probability that the corresponding user belongs to one of the two partitions. The hyper--parameter of METIS, that is the relative size of the partitions, was  tuned by maximizing the number of users whose leaning score is in the 95\% confidence interval between 0 and 1. 
Finally, the tweets of the 10\% of users characterized by the most extreme scores have been read in order to classify them manually as NoVax or ProVax. Every remaining user is assigned to the same partition of the ``extreme'' user with the closest score. 
Using the resulting communities of users, our aim is to monitor the change in the relative size of the two partitions and to measure the polarization over time, where the polarization is the relative density of the in--group agreement with respect to the out--group agreement. Following \cite{chen2021polarization}, we calculated, for each timestamp $t$, the polarization score $S_{t}$ as:
\begin{equation}
S_t=\frac{(E_n+E_p-E_o)}{(E_n+E_p+E_o)}
\label{eq:pol}
\end{equation}
where $E_n$ is the edge density in the NoVax partition --- calculated as the number of observed edges within the community divided by the total amount of the possible edges in the NoVax partition, i.e., $|\mathcal{V}_n| \times (|\mathcal{V}_n|-1)/2$ ---, $E_p$ is the edge density in the ProVax partition and, finally, $E_o$ is the density of edges connecting the two partitions --- calculated as the number of edges linking NoVax and ProVax users normalized by the total amount of possible edges between the two communities, i.e. $\mathcal{V}_n\times \mathcal{V}_p$. The closer $S_t$ is to $1$, the denser the in--partition edges are compared with the out--partition edges and the more polarized is the debate. Conversely, a polarization score near $-1$ indicates that the ties of the network are equally distributed within and between communities, meaning that the discussion is not significantly polarized.

\subsection{Vaccination debate analysis through the global multiplexity matrix}
 In order to provide a view of the evolution of the communities in time, we constructed the global multiplexity matrix $M$, which has proven to be successful in dynamical network analysis \cite{azevedo2021multilayer,hristova2016international}.
In our case, both rows and columns of $M$ correspond to active users, while each entry $M(i,j)$ counts the number of times that users $i$ and $j$ belong to the same community in the forty--one periods considered in our analysis. In fact, the monthly networks would not really represent a multiplex. Multiplex networks \cite{dimitri2017multiplex,battiston2017new,battiston2014structural} are, indeed, made up of the same set of nodes over time. Therefore, we decided to construct $M$ based on all the users who participated in the discussion at any moment during the considered period of time --- which is not an issue since our analysis aimed to select only those nodes that always belong to the same community, during the whole period.
 This revealed core NoVax users, as well as core ProVax users over time.\\ 
Finally, we have identified whether the core communities (for NoVaxes and ProVaxes) were composed by \textit{verified} users or not, also counting the number of connections of the corresponding nodes.
Verified users --- which in Twitter are indicated with a blue tick --- are people (journalists, actors, conductors, singers, etc.) or companies whose identity has been checked by Twitter, guaranteeing the profile authenticity. Therefore, identifying if the most active and influential users, from both sides of the debate, belong to this category represents an important issue. The number of verified users' followers is also a fundamental indicator to quantify how much they are able to spread information. 

\section{Results and discussion}
\label{sec:res}
\subsection{Temporal evolution of the network structure}
\begin{figure*}[h!]
\centering
\includegraphics[width=3.7in]{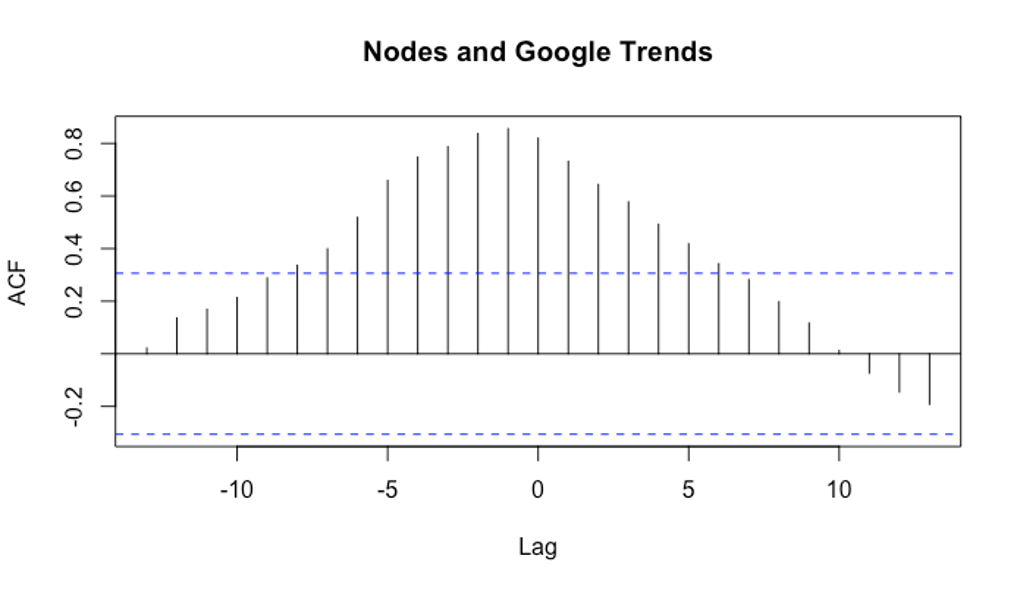}%
\caption{Cross-Correlation between the time-series of the number of nodes and the time-series of the research interest on Google Trend.}
\label{fig:cross}
\end{figure*}

\begin{figure*}[!t]
\centering
\includegraphics[width=3.4in]{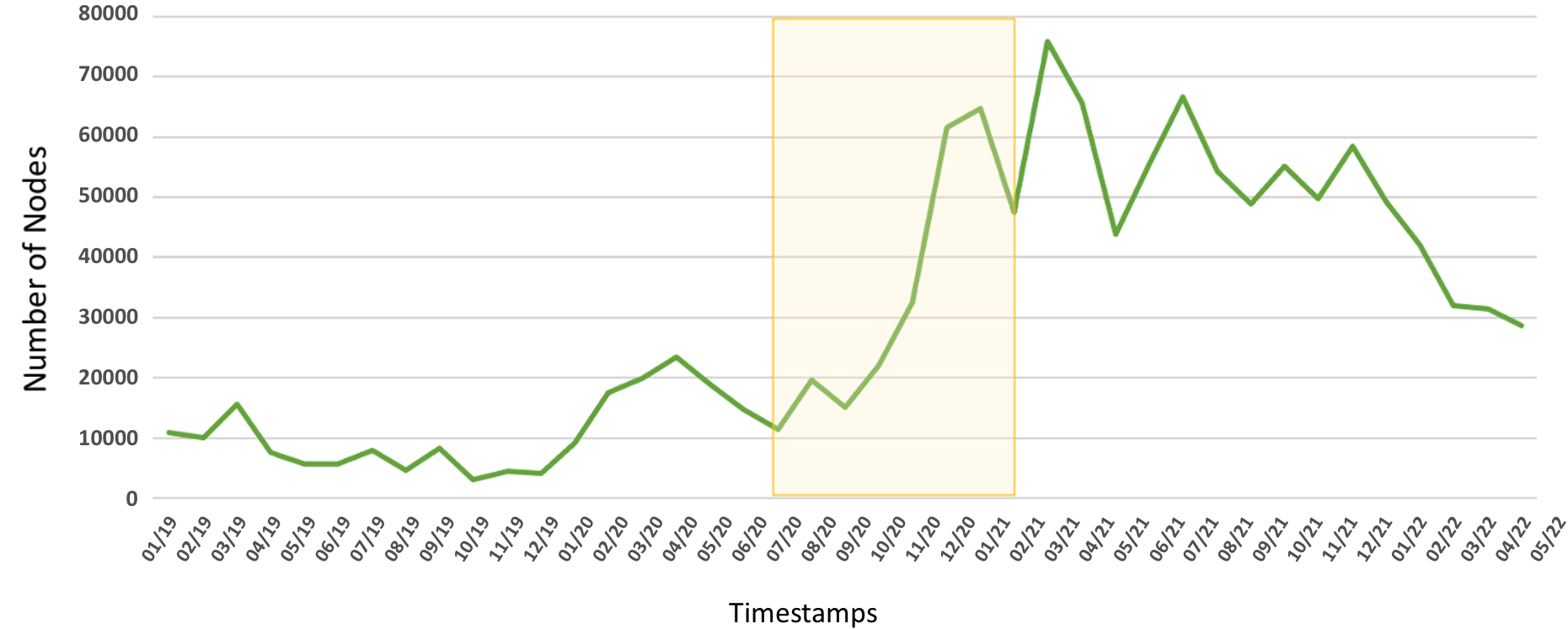}
\hfil
\includegraphics[width=3.4in]{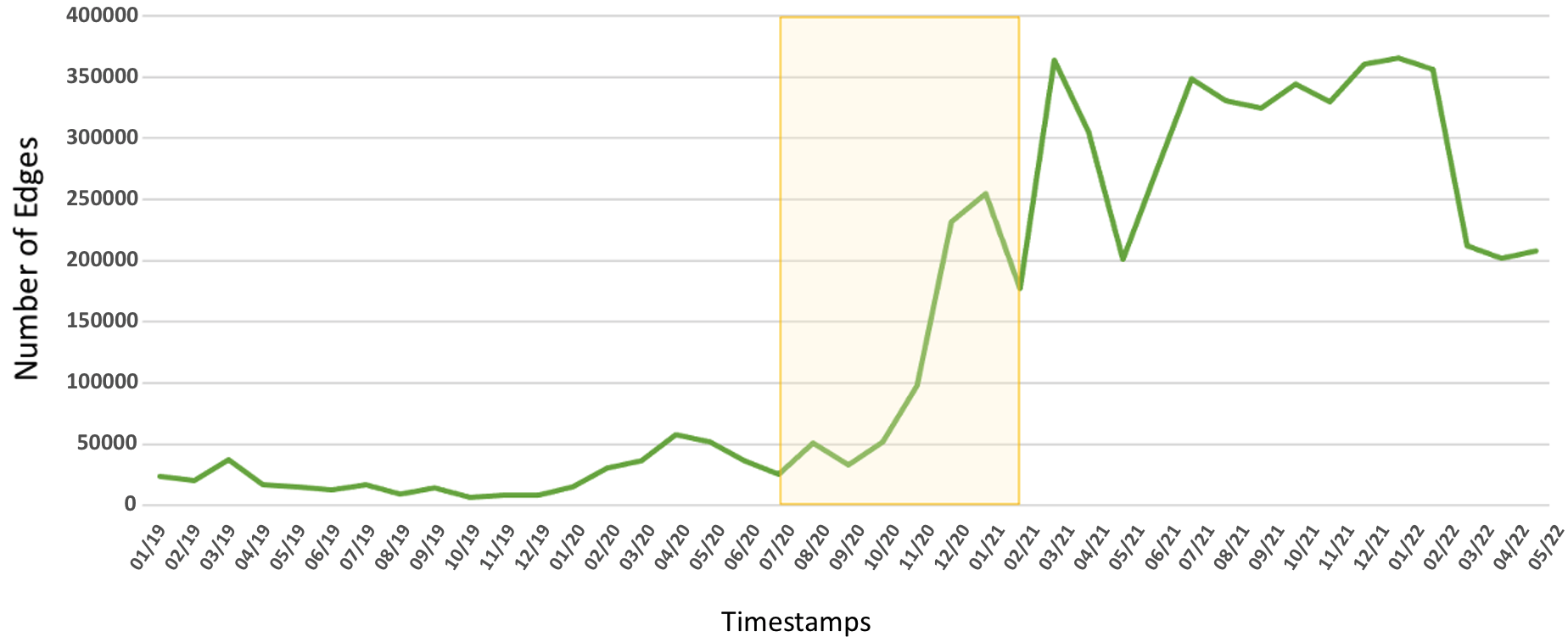}
\caption{Evolution of the number of nodes (left) and edges (right) of the retweet network in time (during the forty--one months considered). The plots clearly show how the beginning of the second wave of COVID-19, which caused the second lockdown and the highest number of deaths in Italy, has changed the dimension of the debate significantly.}
\label{fig:edges}
\end{figure*}

The retweeting relationship between users allowed us to construct forty--one retweet networks, one for each month. The giant connected component present in all the monthly networks includes more than 90\% of the respective total number of nodes. As can be deduced from Figure \ref{fig:edges}, both nodes and edges follow an increasing trend over time until the beginning of 2022, witnessing a huge growth of people interacting on vaccines. Such an increase in the volume of the discussion, after the COVID-19 outbreak, was easily foreseeable. In particular, a real soar of users and exchange of tweets was registered around the 21st month considered in our analysis. This is not surprising, as October 2020 coincides with the onset of the second and more aggressive wave of COVID-19 and, in turn, the spread of rumors about the imminent release of anti--Covid vaccines. In fact, after a few months during which the debate was weaker (see also Figure \ref{fig:heat}, where the bottleneck in the clustering leaning score is visible), a strong peak can be observed in the number of nodes and edges in early autumn 2020. After that, the volume of the debate seems to remain high, until a final declining phase around April--May 2022, due to the pandemic control with the consequent decrease of interest on the topic. Interestingly, we detected a positive cross--correlation (with lag equal to $-1$) between the time series of the Twitter users engaged in vaccine--related discussions and the number of Google searches, based on the same keywords, obtained from Google Trend (Figure \ref{fig:cross}).

\begin{figure*}[!t]
\centering
\includegraphics[width=3.5in]{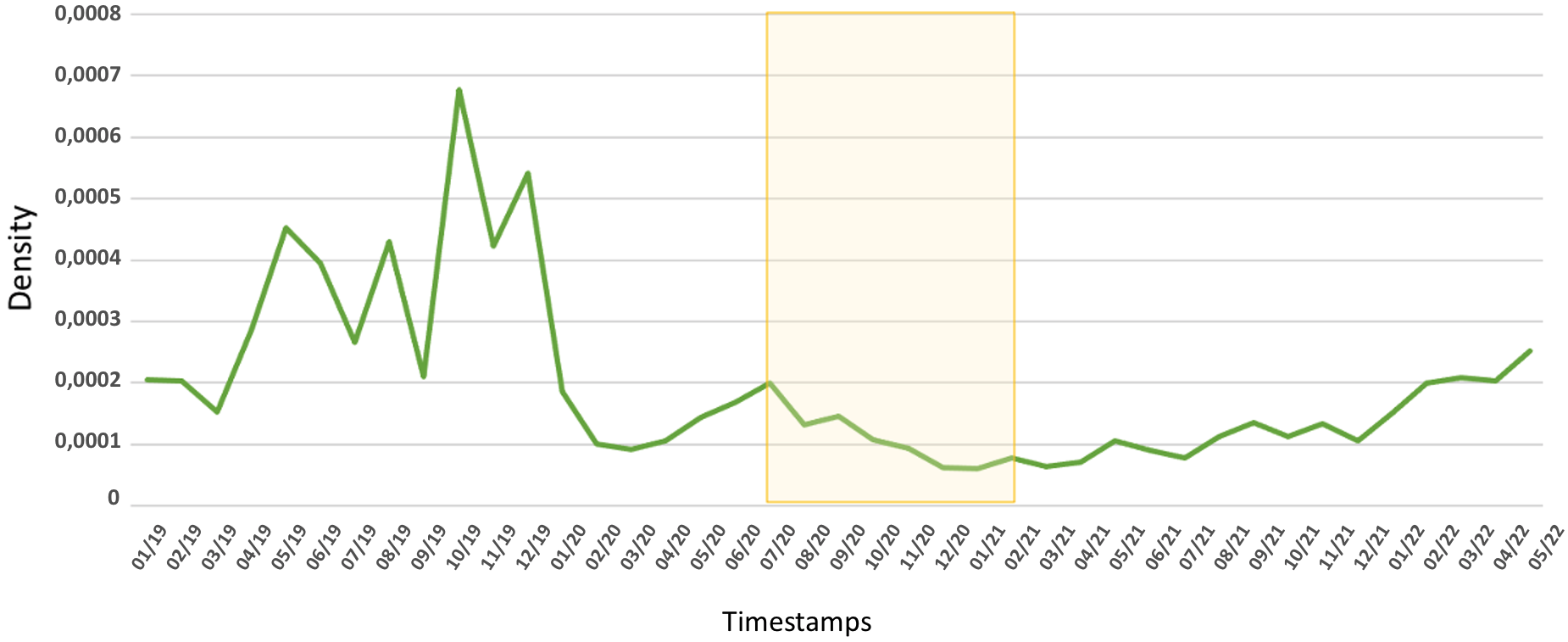}
\hfil
\includegraphics[width=3.5in]{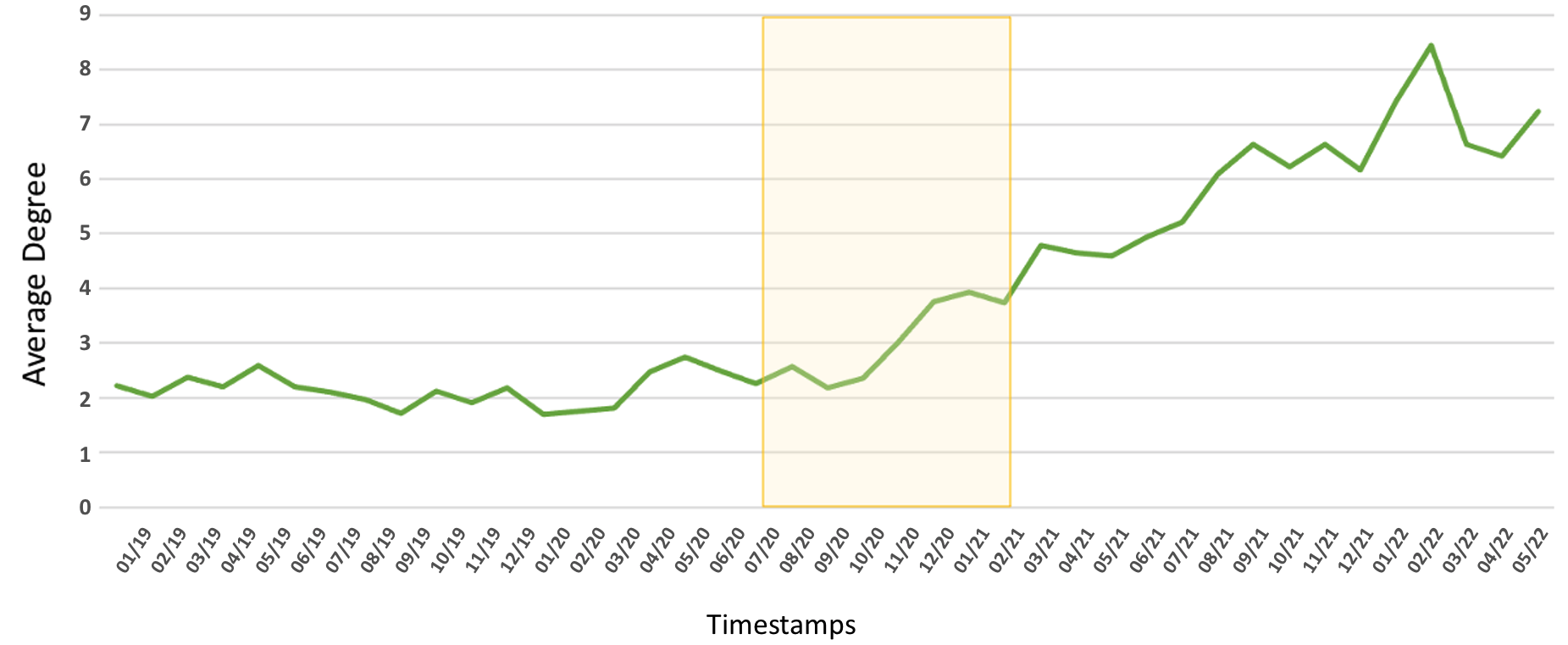}
\includegraphics[width=3.53in]{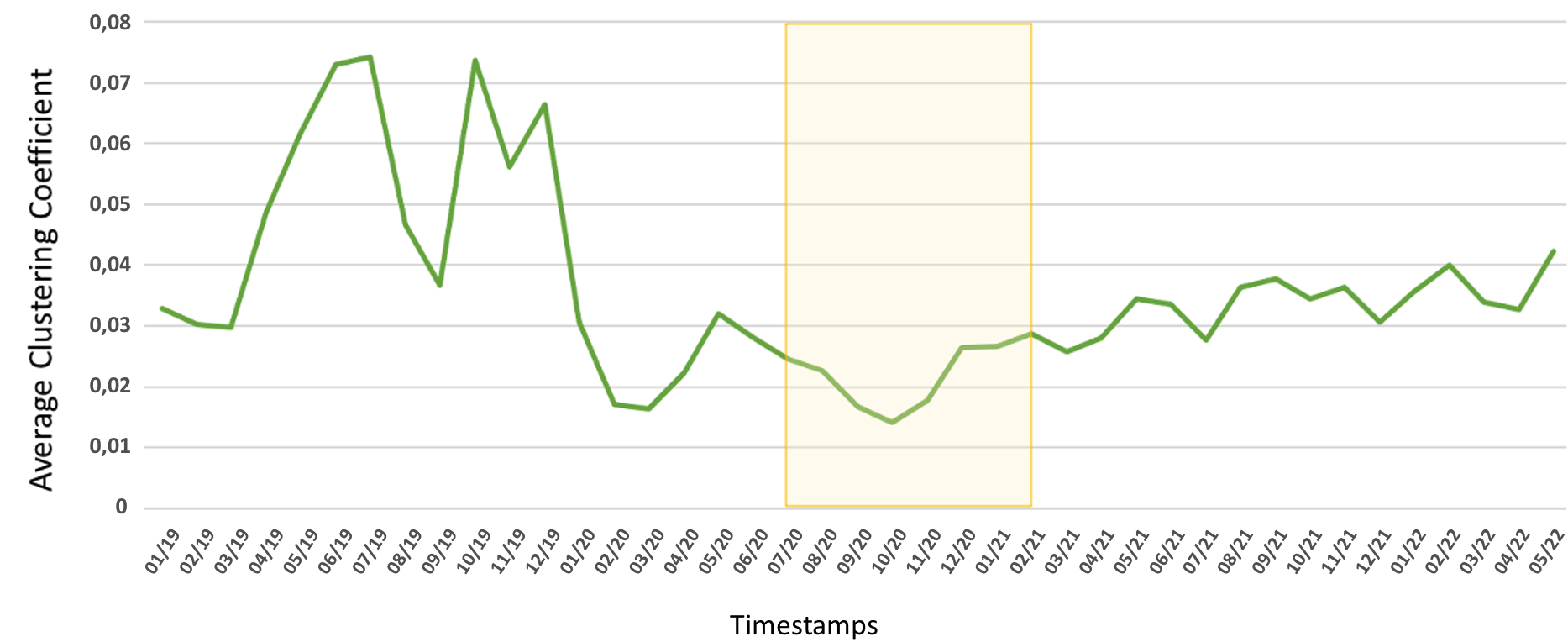}
\hfil
\includegraphics[width=3.5in]{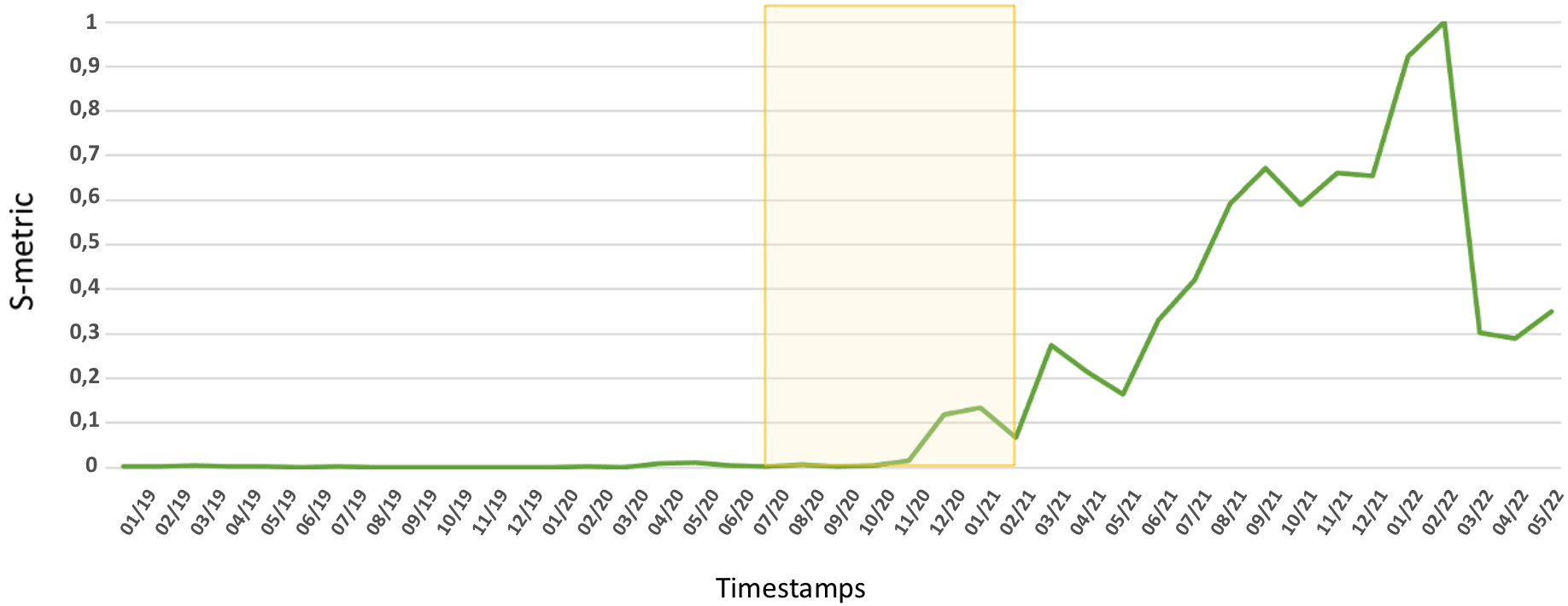}
\hfil
\caption{Evolution of network metrics over time. In all the plots, the $x$ axis shows the month numbers, while the $y$ axis describes the values of the four network measures considered (density, average clustering coefficient, average degree and s--metric). The yellow box indicates the transition period, reported also in Figure \ref{fig:edges}, approximately ranging from July 2020 to February 2021.}
\label{fig:metric}
\end{figure*}

\begin{figure*}[h!]
\centering
\includegraphics[width=5.7in]{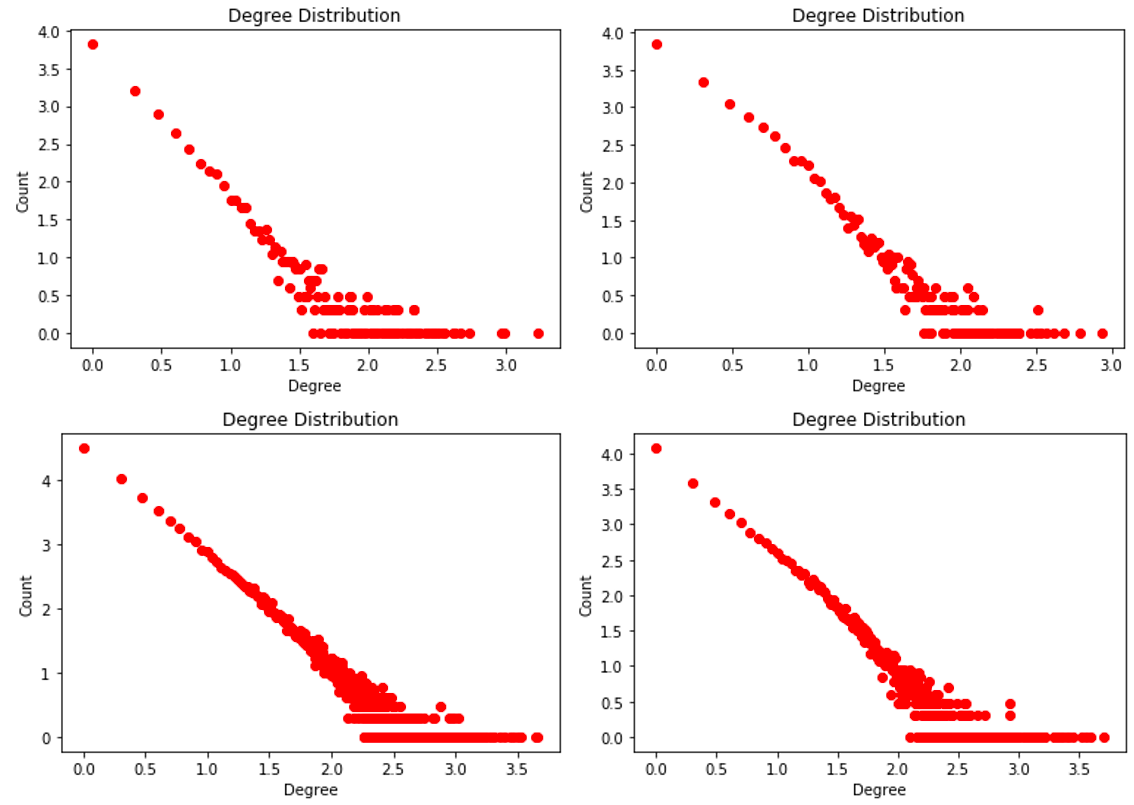}%
\label{fig_second}
\caption{Degree distribution of the retweet networks at months 1, 21, 31, 41. All the networks are scale--free.}
\label{fig:degree}
\end{figure*}

\noindent 
However, it is interesting to monitor how the structure of the debate has evolved over time. For this purpose, we computed, for each monthly network, some structural metrics: density, mean clustering coefficient, mean node--degree and s--metric. Their values over time are plotted in Figure \ref{fig:metric}. Retweet networks show a decrease in the density metric through early 2022 (Figure \ref{fig:metric}, top left). 
Thus, even if there is a growth over time in both the number of nodes and edges, they do not have the same decreasing rate: as the number of users grows, the connection between them becomes more sparse, making the network less similar to a complete graph. This tendency is confirmed also by the negative trend of the average clustering coefficient (Figure \ref{fig:metric}, top right), which implies a decrease in the probability of detecting triangle patterns within the network. Such findings, together with the rise of the average node--degree (Figure \ref{fig:metric}, bottom left), lead to conclude that the structure of the debate significantly changed over time. In particular, although all the forty--one networks show a scale--free pattern --- as revealed by the power--law degree distribution (Figure \ref{fig:degree}) and by the low clustering coefficient \cite{wu2008community} ---,  at the beginning, the discussion looks more uniformly spread all over the users, and later on it gets more and more concentrated in the hands of a small number of influential nodes. These \textit{hubs} play a key role in the diffusion of opinions about vaccines. More specifically, hubs are nodes characterized by a very high node--degree and most other users tend to retweet them massively, building poor connections with low-degree nodes. The increasing s--metric score (Figure \ref{fig:metric}, bottom right) confirms that the network becomes more scale--free as months go by. \\
Opposite conclusions can be drawn for the structural changes of the debate since the beginning of 2022: the end of the vaccination campaign and the spread of the less aggressive Omicron variant seem to turn off the debate (Figure \ref{fig:edges}) and to make it more similar --- in terms of structure --- to the before--Covid era.\\
Monitoring the evolution of the network structure over time is crucial, as changes in the structure can significantly impact the spread of information and disinformation \cite{gleeson2016effects}. Indeed, such changes lead to a more efficient spread of information, since a sparser network with reduced node clusters allows for faster and wider dissemination of information. This is because the information can travel quickly across the network without getting stuck in highly clustered regions or bottlenecks.
However, this also means that disinformation can spread more efficiently as well, especially if the malicious actors responsible for spam or fallacious content have managed to infiltrate the network's key nodes. Disinformation campaigns can take advantage of a sparse network structure to target a wider audience, and the reduced clustering can make it difficult for fact--checkers and debunkers to combat the spread of false information.

\subsection{Temporal evolution of the community composition and network polarization}
\begin{figure}[h!]
\begin{center}
\includegraphics[scale=0.18]{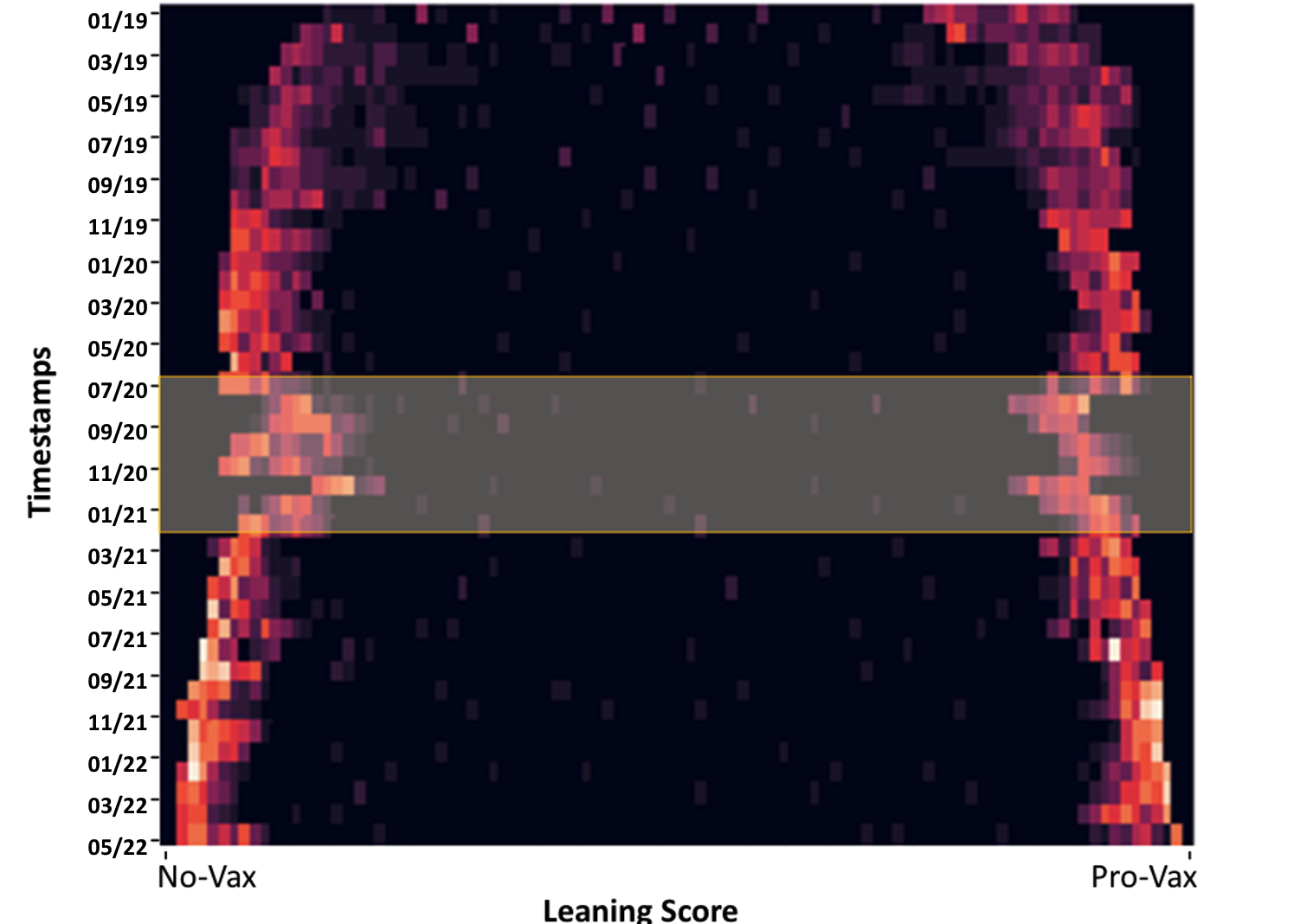}
\end{center}
\caption{Heat--map of the leaning scores. The $x$ axis represents the leaning score value (which varies in [0,1]), while the $y$ axis shows the numbering of the months. The lighter the color the greater the percentage of users sharing the same leaning score.}
\label{fig:heat}
\end{figure}

As previously explained, we applied the METIS algorithm one hundred times to produce two partitions on the forty--one retweet networks. The result of the procedure is presented in the heat--map of the leaning scores shown in Figure \ref{fig:heat}. Each row of the heat--map corresponds to a month (between 1 and 41) while each column is a number between 0 and 1 (represented with at most two decimal places), defining a possible leaning score.  Indeed, the leaning score (see Section \ref{subsec:communityDetection}) describes the probability that a user belongs to one of the two communities. The color of each entry represents the fraction of users that, in the corresponding sub--period, share that leaning score: the lighter the color the higher the percentage of users characterized by that score. Intuitively, a huge fraction of users assigned to one of the two extremes witnesses a highly partitioned network. Based on Figure \ref{fig:heat}, as the months go by, there is a higher percentage of users with extreme leaning scores. This suggests that the network structures become increasingly partitioned, enabling the algorithm to more accurately detect communities. However, a bottleneck is present approximately in correspondence of August 2020--February 2021. 
In this period, a new lockdown started, during which each region was assigned a color corresponding to the spread of the virus and consequently to the severity of the restrictions imposed on the population. This situation corresponds to a highly uncertain time frame and is coherent with the abrupt increase in the debate size, shown in Figures \ref{fig:edges} and \ref{fig:metric} (highlighted by the yellow box). Another plausible explanation for this bottleneck could be the concurrent rollout of the vaccination campaign, which garnered heightened public attention and may have acted as a connecting factor between the two communities during that particular period.
After the bottleneck, a continuous increase of the clustering leaning score is observed, indicating a stronger polarization of the debate.

\begin{figure}[htp]
\begin{center}
\includegraphics[scale=0.4]{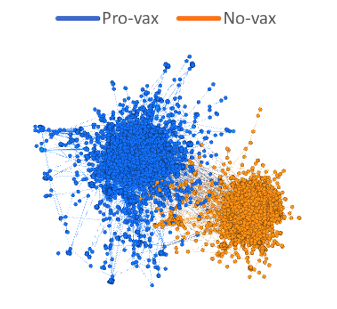}\\
\includegraphics[scale=0.3]{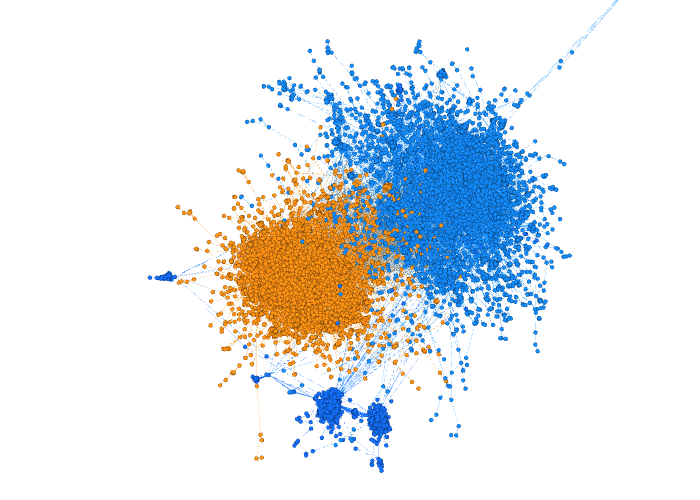}\\
\includegraphics[scale=0.2]{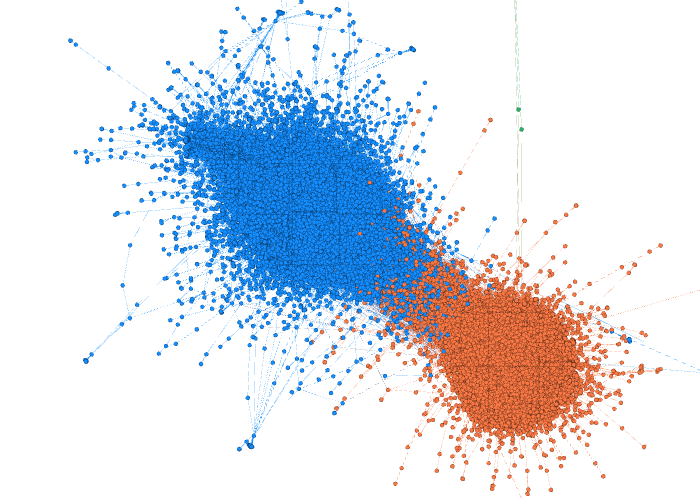}\\
\includegraphics[scale=0.2]{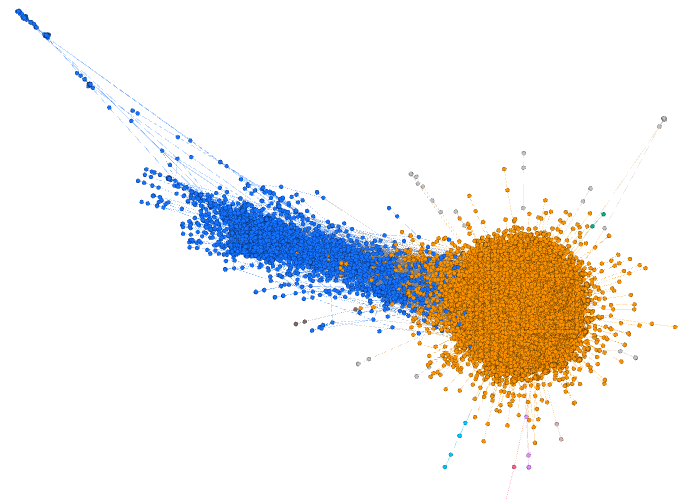}
\end{center}
\caption{Graph representation of the retweet networks at months 1, 21, 31, 41; the blue and orange nodes represent the ProVax and NoVax users, respectively.}
\label{fig:graph}
\end{figure}

\begin{figure*}[h!]
\centering
\includegraphics[width=5.5in]{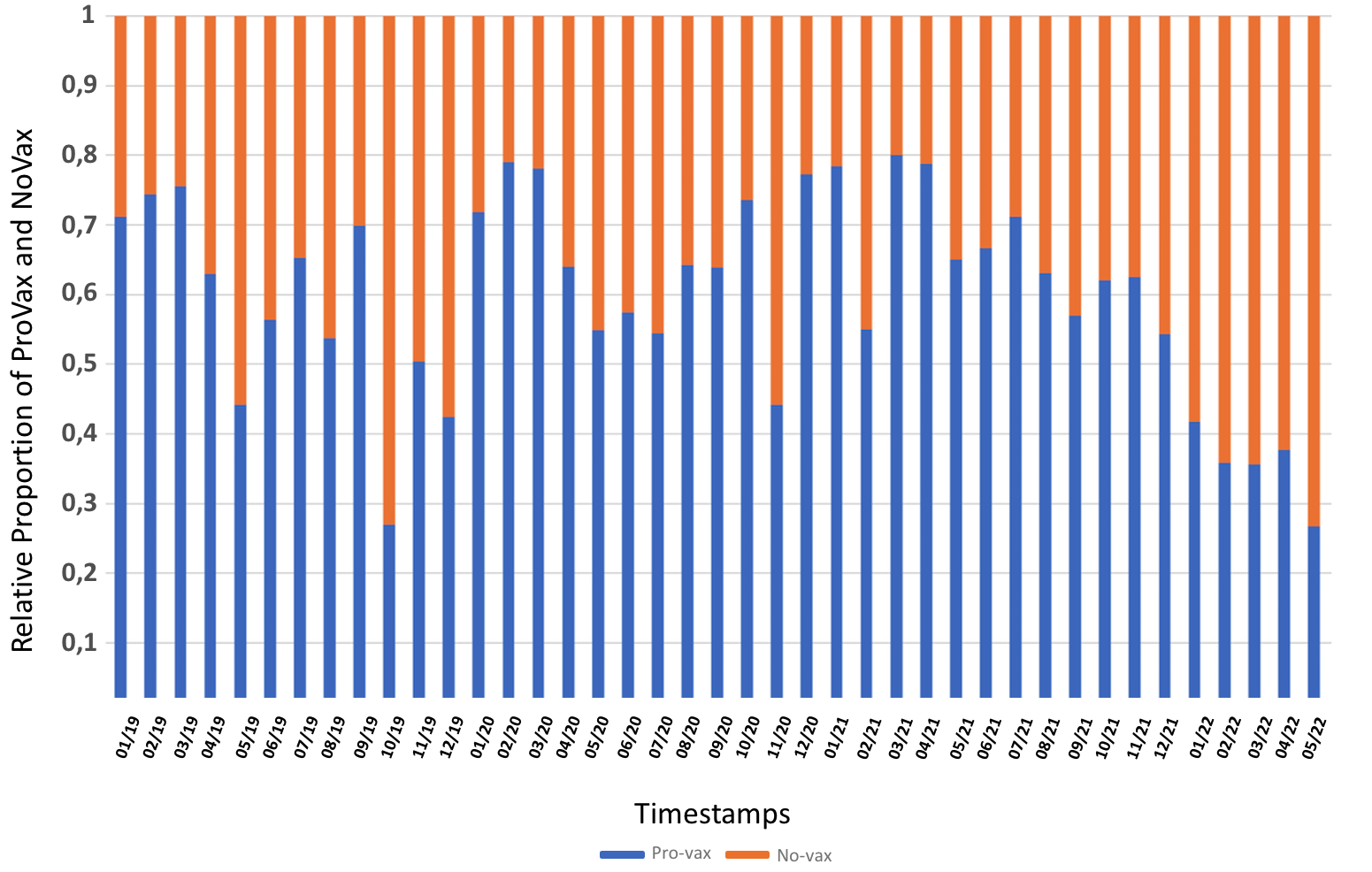}%
\caption{Relative proportion of NoVax (orange) and ProVax (blue) users over the forty--one months, from January 2019 to May 2022.}
\label{fig:proportion}
\end{figure*}

\noindent
In order to characterize the two communities found by the algorithm, the tweets of the users with the most extreme scores were qualitatively read and assessed. For each month, users with similar extreme leaning scores tweeted similar content, guaranteeing the fairness of the partitioning procedure. Every other node of the network is then classified based on the community membership of the ``extreme'' user with the closest score. Four examples of the retweet networks, with the respective communities, are reported in Figure \ref{fig:graph}. From now on the blue and orange colors will indicate ProVax and NoVax users, respectively.
Once the two communities were identified, it was possible to study the relative proportion of vaccination skeptics and advocates for each sub--period (Figure \ref{fig:proportion}). Before the 27th month (March 2021), the relative proportion of ProVax and NoVax users did not follow a particular trend and, mostly, the former were more numerous than the latter. Since April 2021, however, vaccine advocates have become progressively less active, being outnumbered by skeptics from January 2022.

\begin{figure*}[h!]
\begin{center}
\includegraphics[scale=0.25]{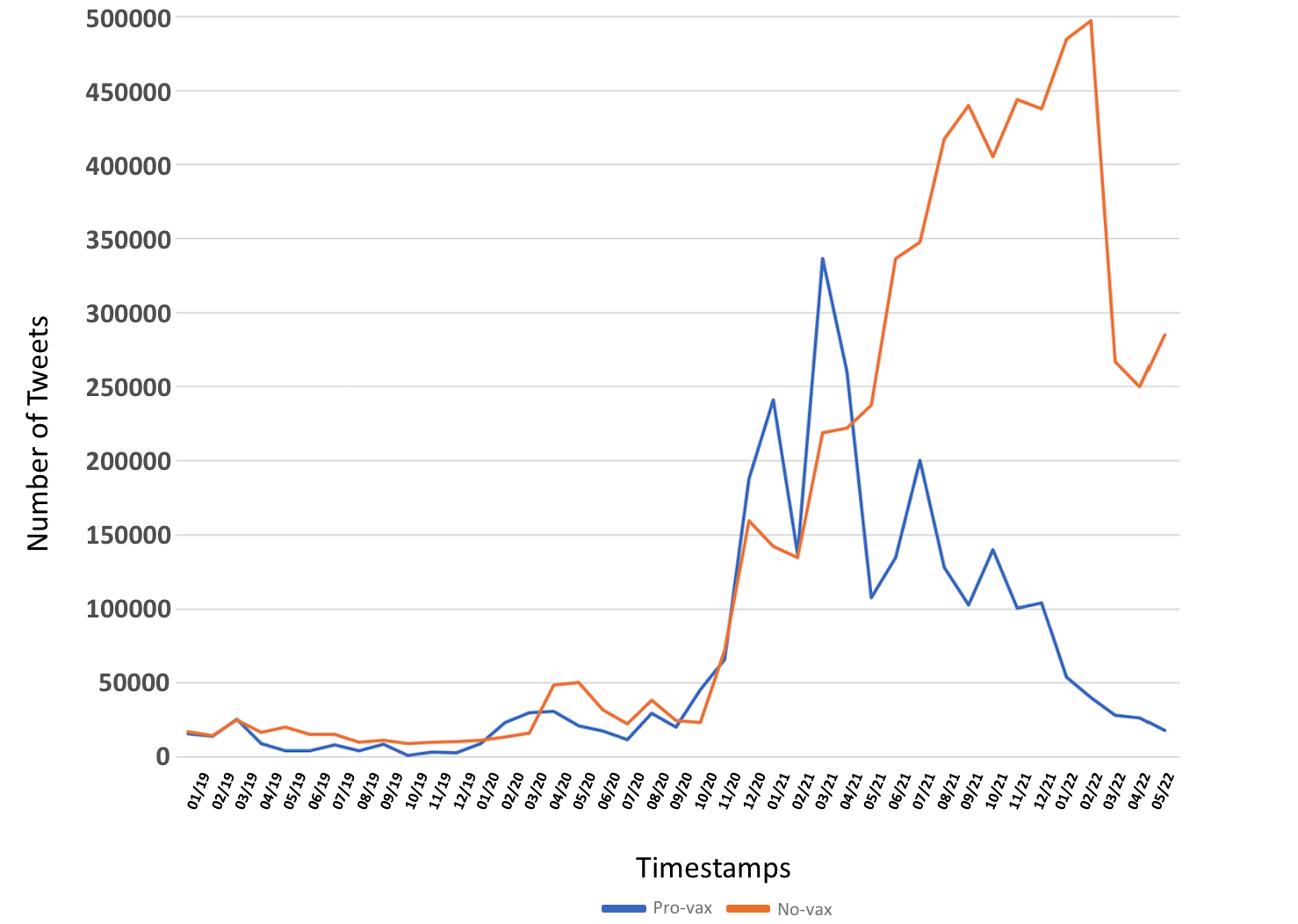}
\end{center}
\caption{Total number of tweets from NoVax (orange) and ProVax (blue) users for each month.}
\label{fig:produc}
\end{figure*}

\begin{figure}[h!]
\begin{center}
\includegraphics[scale=0.17]{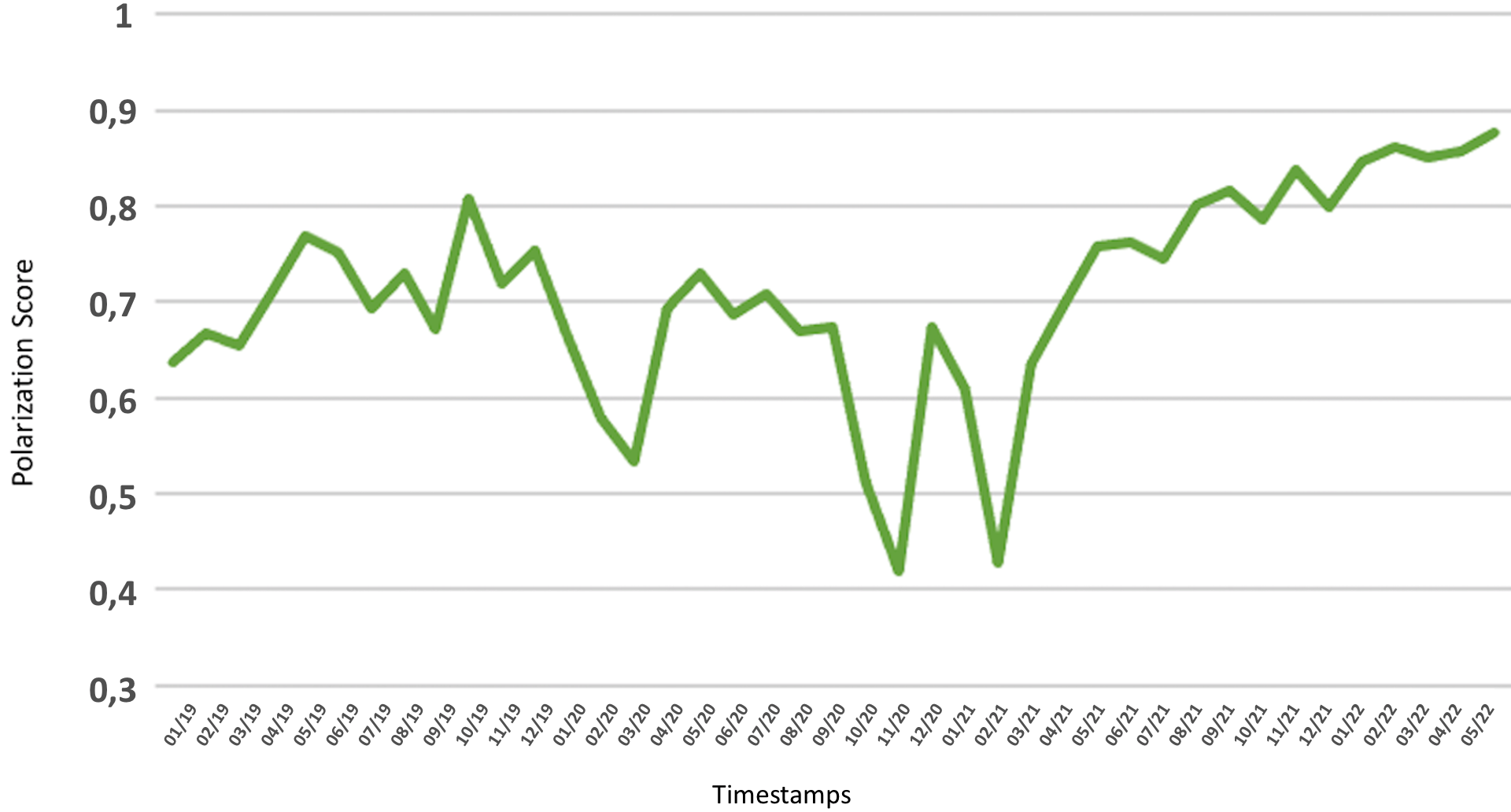}
\end{center}
\caption{Polarization score over the forty--one months considered.}
\label{fig:polariz}
\end{figure}

\noindent
An interesting insight we could gather from our analysis is also given by the number of tweets exchanged within the two communities. As shown in Figure \ref{fig:produc}, NoVax users were generally more productive than ProVax users. The latter tweeted more than the former only during the first five months of the vaccination campaign. Starting from the 28th month (April 2021) the productivity of the two groups begins to follow two different trends: NoVax users become increasingly productive, while the number of tweets from ProVax users decreased significantly.
Furthermore, with the beginning of the COVID-19 vaccination campaign, the subsequent political issues around the green pass, and the introduction of mandatory vaccination law for some categories of workers, the vaccination skeptics became not only more present in the debate, but also more and more active. 
Going into detail, the four peaks of the ProVax curve correspond exactly to the administration of the vaccine to medical personnel (December 2020) and to the first, second and third dose for the entire population (indicatively, April, July and November 2021). In the case of NoVax users, instead, the two most significant peaks relate to the imposition of the green pass obligation (August 6, 2021) and the order of the Ministry of Health (February 8, 2022) for the cessation of the obligation to wear outdoor masks, while the local minimum in April 2022 corresponds to the end of the state of emergency. The correspondence between the peaks of ProVax and NoVax activities and national--level events suggests a substantial difference of base motivation between the two groups, with the NoVax being mostly driven by self--interest considerations.
Amidst the vast body of research on vaccine hesitancy, particularly in the context of the COVID-19 pandemic \cite{troiano2021vaccine,yigit2021evaluation,sallam2021covid,razai2021covid}, some studies have focused on exploring the impact of social trust levels on vaccination rates \cite{roy2022potential,jennings2021lack,faezi2021peoples}. In particular, our findings align with those of a study conducted in Italy \cite{kreps2023resistance}, which examined the relationship between vaccination status, social context, social trust, and adherence to core institutional structures, such as the rule of law and collective commitments. The study revealed that individuals with higher levels of social trust are less likely to remain unvaccinated against COVID-19. Conversely, unvaccinated individuals show less support for honoring other collective commitments unrelated to COVID-19, compared to vaccinated individuals, ceteris paribus. These findings highlight the importance of considering a social contract perspective, alongside the social context, in the study of vaccine hesitancy. As such, they have significant implications for guiding policymakers in developing effective strategies to promote vaccination. Specifically, they suggest that appeals emphasizing individual benefits may be more successful in encouraging vaccination compared to appeals centered around collective responsibility.\\
Finally, the polarization score described in Eq. (\ref{eq:pol}) was computed for each sub--period (Figure \ref{fig:polariz}). The polarization has followed an increasing trend over time with an oscillating phase corresponding to late 2020 and early 2021. This  confirmed that in the month of the so--called ``Vax--day'' (December 2020) not only the network is less clearly divided but the debate is also less polarized. These observations provide further evidence supporting the assertions made in \cite{crupi2022echoes}, namely that the arrival of COVID-19 has failed to alleviate the echo chamber effect in vaccine discussion. On the contrary, our analysis indicates that the advent of COVID-19 has exacerbated the polarization of the debate.  
\subsection{Possible strategies to slow down vaccine misinformation}
\begin{figure*}[h!]
\begin{center}
\includegraphics[scale=0.2]{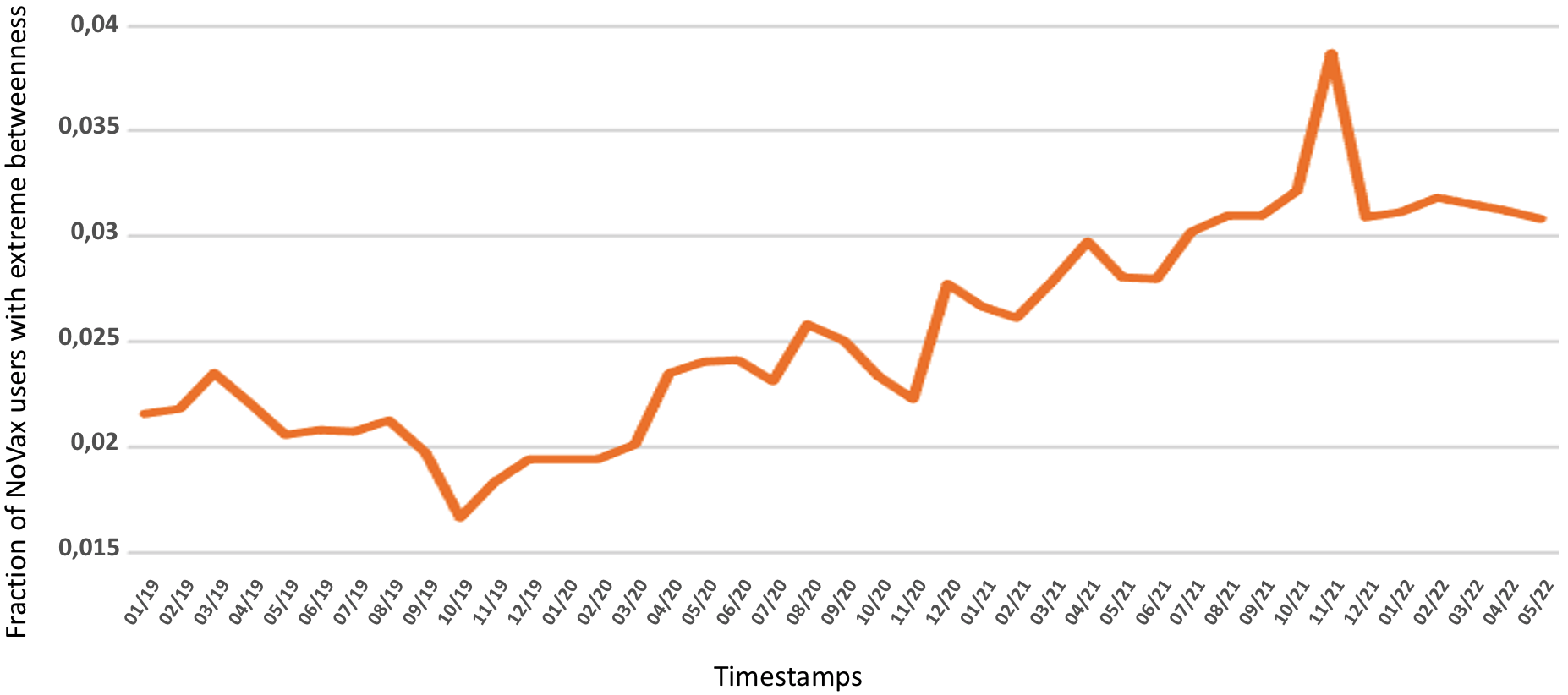}
\end{center}
\caption{Fraction of NoVax users characterized by a betweenness centrality outside the 95th percentile.}
\label{fig:bet}
\end{figure*}

By identifying communities over time, we can develop effective strategies to curb the spread of vaccine misinformation and limit the growth of the NoVax community. One promising approach is to pinpoint nodes with a central role in the NoVax community and disrupt their influence through node attacks, such as account blocking. To determine node importance, we employed the betweenness centrality metric for NoVax nodes. Analyzing the betweenness of NoVax nodes enabled us to examine the pathways of misinformation spreading. Specifically, we analyzed the subgraph induced by the NoVax nodes at each time step and calculated the distribution of betweenness among the nodes in the subgraph. Our analysis revealed that all the distributions followed a power--law trend, indicating that only few nodes had high betweenness at each time step. We then determined the number of nodes with extremely high betweenness by calculating the percentage of nodes outside the 95th percentile of the distribution for each time step (Figure \ref{fig:bet}). Although the number of nodes with high betweenness increased over time, they constituted a small portion of the total nodes. Therefore, account blocking could be an effective strategy to hinder their ability to spread misinformation. Other node attacks, based on metrics such as degree and closeness, could also fragment the NoVax community.
\subsection{Characterization of the core NoVax/ProVax users}
\begin{figure*}[!t]
\centering
{\includegraphics[width=2.2in]{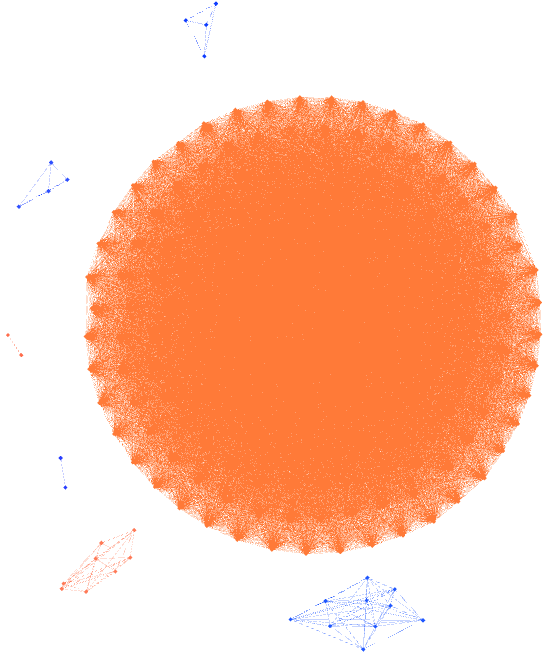}%
\label{fig:g41}}
\hfil
{\includegraphics[width=2.in]{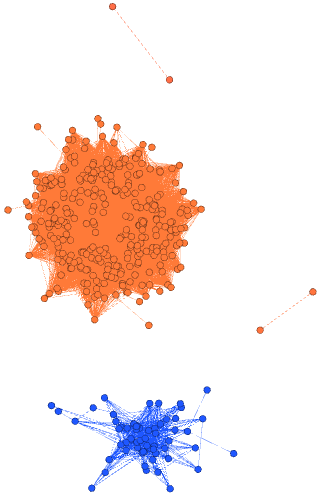}%
\label{fig:g40}}
\hfil
{\includegraphics[angle=90, width=1.7in]{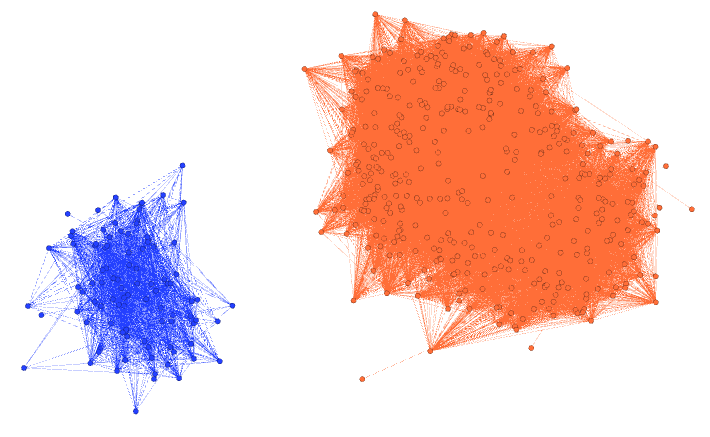}
\label{fig:g39}}
\caption{Users that belong to the same community for forty--one months, forty months and thirty--nine months. The two colors (blue/orange) represent the two core communities (ProVax/NoVax) identified among these subgroups.}
\label{fig_sim}
\end{figure*}

The global multiplexity matrix $M$, of dimension 300,653$\times$300,653, is found to be sparse. Each entry $M(i,j)$ is an integer between 0 and 41 representing the number of months in which the two users $i$ and $j$ belong to the same community. First, we extracted all pairs of users who have been in the same community throughout the period, i.e. pairs of users whose relative matrix entry equals 41. Then, we used the extracted information to construct a new network connecting only such selected users. The resulting graph (Figure \ref{fig:g41}) has seven connected components, whose dimensions are shown in Table \ref{fig:t41}.


\begin{table}
\begin{center}
\caption{Dimension and composition of the seven connected components composed by users who belong to the same community for forty--one months; the percentage is calculated as the relative number of times that the nodes inside the connected components represent NoVax/ProVax users }
\begin{tabular}{|l|c|r|}
\hline
\textbf{\# Nodes in Connected Component} & NoVax & ProVax \\
\hline
183 & \textbf{100\%} & 0\%\\
\hline
10 & 2.4\% & \textbf{97.6\%}\\
\hline
8 & \textbf{95.1\%} & 4.9\%\\
\hline
4 & 4.8\% & \textbf{95.2\%}\\
\hline
4 & 4.9\% & \textbf{95.1\%}\\
\hline
2 & 5.1\% & \textbf{94.9\%}\\
\hline
2 & \textbf{97.5\%} & 2.5\%\\
\hline
\end{tabular}
\label{fig:t41}
\end{center}
\end{table}

\noindent
It is worth noting that each component is fully connected because requiring that nodes belong to the same community forty--one times --- which is the maximum possible --- is a transitive membership relation. As we can see in Table \ref{fig:t41}, the largest connected component is totally composed of users that have always been NoVax, whereas ProVax convinced users populate the second largest component. The other five connected components are composed by few users that have changed their mind about vaccines at least one time during the forty--one months. However, it is also possible that some convinced and active users have not tweeted during some periods. Considering this, we decided to relax the constraint by requiring that nodes belong to the same community just forty times. In this case, the graph (Figure \ref{fig:g40}) has four connected components --- no more fully--connected ---, with the largest one composed by the core NoVax users (Table \ref{fig:t40}).  

\begin{table}
\begin{center}
\caption{Dimension and composition of the four connected components composed by users who belong to the same community for forty months }
\begin{tabular}{|l|c|r|}
\hline
\# Nodes in Connected Component & NoVax & ProVax \\
\hline
312 & \textbf{96.5\%} & 3.5\%\\
\hline
60 & 3.2\% & \textbf{96.8\%}\\
\hline
2 & \textbf{95.1\%} & 4.9\%\\
\hline
2 & \textbf{95\%} & 5\% \\
\hline
\end{tabular}
\label{fig:t40}
\end{center}
\end{table}

\noindent
Decreasing the number of periods to thirty--nine, the two core NoVax and ProVax sets became larger and more stable: the graph at this point contains only two connected components (Figure \ref{fig:g39}), whose dimension and composition are shown in Table \ref{fig:t39}. Based on Table \ref{fig:t39}, we can also conclude that the NoVax supporters are more numerous and stable over time, whereas the core set of ProVax users is significantly smaller. Anyway,  the most convinced and active ProVax users have 100200 followers on average, while NoVaxes have generally less followers (44067), even if they are more constant and more present in the debate. Furthermore, none of the users who make up the NoVax backbone are verified users, while 15.85\% of vaccine advocates are. \\
Another noteworthy observation pertains to the structure of the core NoVax and ProVax users. As depicted in Figures 10b and 10c, it is evident that the NoVax users exhibit a significantly more homogeneous structure compared to the ProVax users. This characteristic may be closely tied to their efficacy in disseminating false information. As highlighted in \cite{pierri2022online}, the act of sharing false information, even if subsequently debunked, tends to correlate with individuals either postponing vaccination or outright rejecting it. 

\begin{table}
\begin{center}
\caption{Dimension and composition of the two connected components composed by users who belong to the same community for thirty--nine months}
\begin{tabular}{|l|c|r|}
\hline
\textbf{\# Nodes in Connected Component} & NoVax & ProVax \\
\hline
405 & \textbf{95.6\%} & 4.4\%\\
\hline
97 & 3.3\% & \textbf{96.7\%}\\
\hline
\end{tabular}
\label{fig:t39}
\end{center}
\end{table}

\begin{figure*}[h!]
\begin{center}
\includegraphics[scale=0.25]{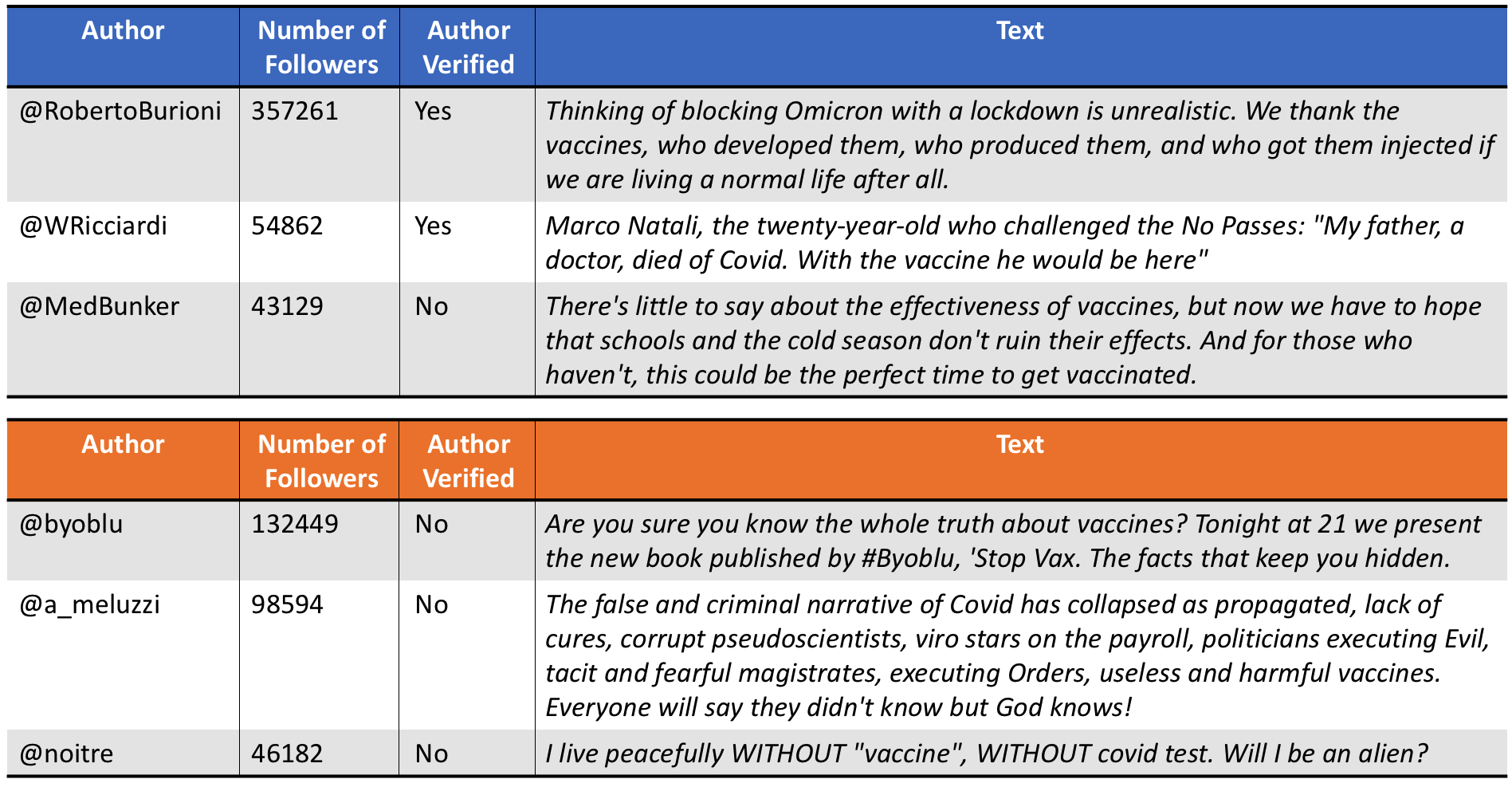}
\end{center}
\caption{Selection of the three most followed users in the ProVax (blue) and NoVax (orange) core sets. The texts are translated from Italian to English using Google Translate.}
\label{fig:text}
\end{figure*}

\noindent
Figure \ref{fig:text} shows a selection of the most popular users on both
sides. Interestingly, among ProVaxes, the most followed user is Dr. Roberto Burioni, a famous Italian virologist who played an important role in health information during the COVID-19 pandemic and even before. The most influential and active NoVax user is instead Byoblu
, a counter--information blog founded by Claudio Messora, a well--known member of the NoVax community in Italy, as well as an advocate of conspiracy theories on the health dictatorship.

\section{Conclusion}
\label{sec:concl}
In Italy, the debate on vaccines has always been very heated and COVID-19 has contributed to exacerbating it. Therefore, the discussion triggered by the beginning of the pandemic on social networks provides an accurate picture of the perception of Italians on this burning issue. \\ 
In order to study the temporal evolution of the vaccination debate during the last three years, we collected the new dataset TwitterVax, using the Twitter API and crawling tweets related to several vaccine keywords. 
This first collection step allowed us to exploit Twitter data from a network perspective, in order to understand the changes in the structure, scale and polarization of the vaccine dispute in Italy, starting from January 2019 to May 2022. Based on the collected data, we showed that, with the outbreak of COVID-19, the debate has not only dramatically intensified, but has also become more concentrated in the hands of a few influential hubs, who have played a vital role in disseminating vaccine information. However, this trend seems to have reversed since the beginning of 2022, testifying that the change in the structure of the discussion was probably due to the COVID-19 emergency. We successfully detected the NoVax and ProVax communities, demonstrating that the relative proportion of the two user cohorts does not change significantly over time. In particular, the NoVax community is often the least populated but also the most active in terms of the number of tweets. Moreover, we computed a polarization score between the two user groups, demonstrating an increasing level of polarization. Finally, using a multiplexity approach, we identified the core of NoVax and ProVax users. In this way, we have verified that core NoVaxes are more than core ProVaxes who, however, have more followers on average and a higher percentage of verified users.\\
Future perspectives include the possibility to collect a more complete data set, covering a larger time period, as well as process the data according to a higher resolution. It could also be interesting to analyze higher order motifs in temporal multiplex networks, which could give us further information on the structure of the interaction networks \cite{li2023multiplex,boccaletti2014structure}.
Thereby, other strategic events that have had a profound impact on the debate can be identified and analyzed, also using different network approaches and evaluation metrics. Furthermore, our analysis did not consider the presence of artificial robots (bots) in the vaccine debate on Twitter. Identifying and filtering bot users can be a challenging task\cite{alothali2018detecting,ferrara2016rise}, which is beyond the scope of this work. However, analyzing the percentage of bots and investigating their influence on diffusion and polarization processes would be valuable and will be the subject of future research.

\bibliographystyle{unsrt}  
\bibliography{refs1}

\end{document}